\documentclass[conference]{IEEEtran}
\usepackage{cite}
\usepackage{amsmath,amssymb,amsfonts}
\usepackage{bm}
\usepackage{algorithmic}
\usepackage{graphicx}
\usepackage{textcomp}
\usepackage{xcolor}
\usepackage{tikz}
\usepackage{caption}
\usepackage{subcaption}
\usepackage{listings}
\usepackage[frozencache,cachedir=.]{minted}
\def\BibTeX{{\rm B\kern-.05em{\sc i\kern-.025em b}\kern-.08em
    T\kern-.1667em\lower.7ex\hbox{E}\kern-.125emX}}
\newcommand\mytt[1]{\texttt{{#1}}}
\begin{document}
\lstset{language=C++}
\title{ProtoX: A First Look\\
%{\footnotesize \textsuperscript{*}Note: Sub-titles are not captured in Xplore and
%should not be used}
% \thanks{This project is funded by the  DOE Office of Advanced Scientific Computing Research the Office of Science of the U.S. Department of Energy under Contract No. DE-AC02-05CH11231.}
}

\author{\IEEEauthorblockN{Het Mankad\IEEEauthorrefmark{1}, Sanil Rao\IEEEauthorrefmark{1}, Phillip Colella\IEEEauthorrefmark{2}, Brian Van Straalen\IEEEauthorrefmark{2}, Franz Franchetti\IEEEauthorrefmark{1}}
\IEEEauthorblockA{\IEEEauthorrefmark{1}\textit{Electrical and Computer Engineering Department Carnegie Mellon University, Pittsburgh, PA, USA} \\
\IEEEauthorblockA{\IEEEauthorrefmark{2}\textit{Lawrence Berkeley National Laboratory, Berkeley, CA, USA}\\
\IEEEauthorrefmark{1}\{hmankad, sanilr, franzf\}@andrew.cmu.edu, \IEEEauthorrefmark{2}\{bvstraalen, pcolella\}@lbl.gov}}
}

\maketitle

\begin{abstract}
We present a first look at ProtoX, a code generation framework for stencil and pointwise operations that occur frequently in the numerical solution of partial differential equations. ProtoX has Proto as its library frontend and SPIRAL as the backend. Proto is a C++ based domain specific library which optimizes the algorithms used to compute the numerical solution of partial differential equations. Meanwhile, SPIRAL is a code generation system that focuses on generating highly optimized target code. Although the current design layout of Proto and its high level of abstractions provide a user friendly set up, there is still a room for improving it's performance by applying various techniques either at a compiler level or at an algorithmic level. Hence, in this paper we propose adding SPIRAL as the library backend for Proto enabling abstraction fusion, which is usually difficult to perform by any compiler. We demonstrate the construction of ProtoX by considering the 2D Poisson equation as a model problem from Proto. We provide the final generated code for CPU, Multi-core CPU, and GPU as well as some performance numbers for CPU.
\end{abstract}


\section{Introduction}
There is a myriad of application areas in the field of scientific computing and engineering where numerical solutions to partial differential equations (PDEs) are required to be computed. Numerical methods like the finite difference method (FDM), finite element method (FEM), finite volume method (FVM) and multigrid method are used to approximate the solutions to these PDEs. The key components of these algorithms are the stencil and pointwise operations. 
Stencils are defined as a linear transformation,
\begin{align}\label{stencil}
    S(x)_{i} = \Sigma_{j} \alpha_{j}x_{i+j},
\end{align}
where $i, j \in \mathbb{Z}^{D}$ with $D$ denoting the number of space dimensions, $\alpha$ denotes the weight and $x$ is the multidimensional data array.

Typically these numerical algorithms are iterative in nature resulting in performing the stencil operations multiple times.  
Writing codes for these stencil based methods from scratch can be quite a cumbersome task for anyone. 
As such developers turn to libraries that provide stencil operations for them. One such library is Proto. It is a domain specific library written in C++ that provides a high level of abstraction for solving various PDEs using some of the aforementioned numerical methods.

Proto's abstraction enables ease of programmability, but has drawbacks when it comes to performance. Many of Protos' abstractions can be fused and optimized together, resulting in better performance. However, abstraction fusion is something no compiler can easily perform. This results in additional burden on the library developers to manually introduce these optimizations. To enable abstraction fusion in Proto, we propose ProtoX, which is a C++ library based on Proto and runs a code generation system SPIRAL \cite{Spiral2005, Franchetti2005} in the backend. The concept of using SPIRAL in the backend and a C/C++ based library in the front has shown positive results in the past \cite{FFTX, Sanil2020}. Some of the related works in the area of optimizing stencil computation with either automatic code generation or by optimizing data movement involved while performing stencil operation are discussed in the next section. 
 
{\bf{Contribution.}} Some of the main contributions of this work are: 
\begin{enumerate}
    \item A proof of concept of hooking SPIRAL code generation as backend into Proto is presented.
    \item A Proto example is considered as a SPIRAL specification and a full program optimization with merged kernels across C++ functions are discussed for that example.
    \item This work is a first look of ProtoX. Hence, initial CPU results are presented.
\end{enumerate}

This work is organized as follows. Section \ref{sec2} briefly discusses some of the related work in this area of research. Section \ref{background} provides the necessary background material to understand this work. Here we discuss the design layout of both Proto and SPIRAL. Section \ref{sec4} introduces the structure of ProtoX with the help of the 2D Poisson problem and finally section \ref{sec5} and \ref{sec6} provides with the concluding remarks, future work and acknowledgement.

\section{Related Work} \label{sec2}
Due to the frequent use of stencil operations in various scientific applications, there has been a significant amount of work done over the past couple of years in the area of generating optimized code for such computations. Some work focuses on developing stencil codes for a particular application, for example, Halide \cite{halide} is a C++ based DSL focusing on all image processing pipelines. Meanwhile, some focus on platform portability like LIFT\cite{LIFT} and SBLOCK\cite{SBLOCK}. There is also some work done towards improving performance of stencil codes by introducing new ways to optimize the data movement at a compiler level like \cite{BVSCompiler2015}.  PATUS \cite{PATUS} is an example of code generation framework for stencils that provide multi-core CPU and GPU code for such stencil operations. The work done by Holewinski et. al. in \cite{SadayGPU} showcases their efforts in the area of highly optimized code generation for stencils but mainly targeting GPU accelerators. In \cite{ManyCore} Li et. al. work on large scale stencil computation generating code for both spatial and temporal computation on heterogeneous many-core processors. Some of the more recent work in the area of code generation for stencil computation can be seen in ExaStencil \cite{ExaStencil}, YASK \cite{YASK} and Code Generation for In Place Stencils\cite{InPlaceStencil}.

\section{Background}\label{background}
In order to understand the design of ProtoX we will first need to understand the design structure of Proto as well as SPIRAL. In this section we will briefly provide an overview of Proto and SPIRAL.

{\bf Proto.}
Proto is a C++ library designed to provide an intuitive interface that optimizes the designing and scheduling of an algorithm aimed at solving various PDEs numerically. In order to approximate the solution of a PDE, the domain is usually divided into either structured or non-structured grid and the equations are discretized using methods like FDM, FEM or FVM. Currently, Proto takes into account a multidimensional rectangular, structured grid with periodic boundary condition for all model problems.

\begin{figure}
    \centering
    \includegraphics[scale=0.5]{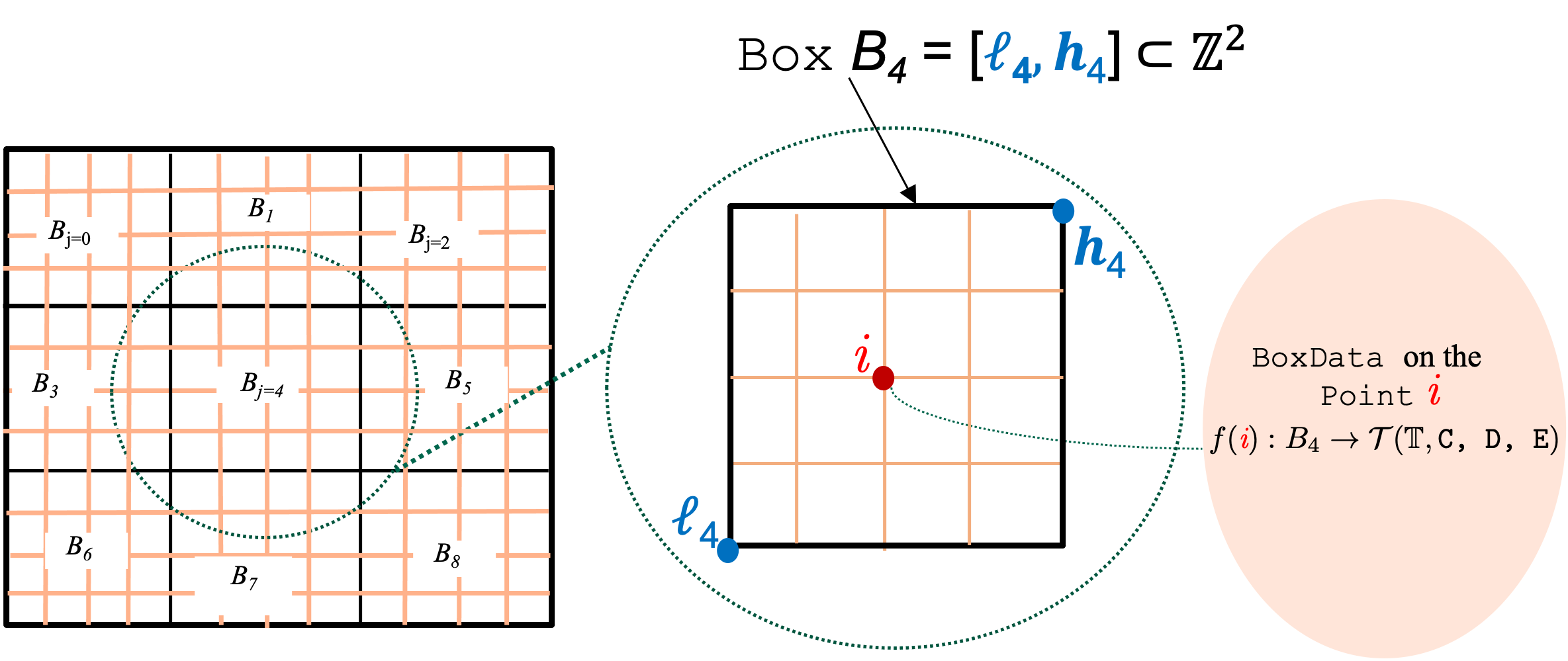}
    \caption{A pictorial representation of the \mytt{Point, Box} and \mytt{BoxData} classes in Proto is shown.}
    \label{ProtoClasses}
\end{figure}

There are four main C++ classes that implement a representation of the spatial rectangular grids are discussed below.
\begin{itemize}
    \item \mytt{Point}: It represents points in $\mathbb{Z}^{D}$. It is used to indicate the location as well as offsets in a grid in Proto.
    \item \mytt{Box}: It denotes the rectangular subsets of $\mathbb{Z}^{D}$ that has elements from the \mytt{Point} class as it nodes. Figure \ref{ProtoClasses} indicates a typical 2D computational domain used in Proto which is divided into several Boxes $B_{j}, j= 0 \dots, M-1, M \in \mathbb{N}$.  Here, each box in the \mytt{Box} class is represented as $B_j = [\boldsymbol{\ell}_j, \boldsymbol{h}_j] \subset \mathbb{Z}^{2}$, where ($\boldsymbol{\ell}_{j} \in \mathbb{Z}^{2}$) and $(\boldsymbol{h}_j \in \mathbb{Z}^{2})$ are it's lowest and highest corner points. 
    \item \mytt{BoxData} $\langle \mathbb{T}, \mytt{C}, \mytt{D}, \mytt{E} \rangle$:  It represents the multidimensional data arrays on each \mytt{Point} in a \mytt{Box}. It is a templated class in C++. For a given point $\boldsymbol{i}$ in $B_j$, the \mytt{BoxData} class is defined as a function,
    \begin{align*}
        f(\boldsymbol{i}): B_j \rightarrow \mathcal{T}(\mathbb{T}, \mytt{C, D, E}).
    \end{align*}
    Here $\mathbb{T}$ is the class template holder for the data which can be either $\mathbb{R, C}$ or $\mathbb{Z}$. Meanwhile, the dimensions of the range space is given by \mytt{C, D} and \mytt{E}. Hence, based on the values of \mytt{C, D} and \mytt{E} one can identify the given data to be a scalar, vector, matrix or $3^{rd}$ order tensor.
    \item \mytt{Stencil}$\langle \mathbb{T} \rangle$: Another important class in Proto is the \mytt{Stencil} class. This is where stencils are defined as self-contained objects to implement the operations described in (\ref{stencil}). Stencils are applied to the \mytt{BoxData} to update the values in there. They are accompanied by the \mytt{Shift} class to carry out the grid shifts that are required in a stencil operation.
\end{itemize}

In order to apply the stencils or any other function in a pointwise manner, Proto uses the \mytt{forall} function. It takes in as an argument the function which is to be applied in a pointwise manner and the corresponding \mytt{BoxData} used for it.  

{\bf SPIRAL.}
SPIRAL is a GAP based code generation system that was initially developed to automatically generate optimized C/C++ programs for linear transforms like the discrete Fourier Transform (DFT), discrete sine/cosine transform and many more \cite{Spiral2005, Franchetti2005}. SPIRAL uses the Signal Processing Language (SPL) \cite{SPL1} to develop algorithms for these signal processing transforms. SPL is a declarative mathematical language that expresses linear transforms as matrix-vector product. Here, the matrix is considered to be an operator with one input and one output vector. 

Operator Language (OL) is a generalization of SPL. It is the result of the ongoing efforts to expand the scope of SPIRAL beyond signal processing transforms \cite{SpiralOL2009}. As described in \cite{SpiralOL2009}, it is a domain specific language (DSL) which is also a mathematical declarative language like the SPL. It is written in terms of different operators that describe the data flow and data layout of a given transform or a computational kernel of interest. This helps in manipulating the algorithm in such a way that enables introducing multiple optimization techniques at a high level of abstraction. Table \ref{operators} provides a list of some of the operations that will be useful in describing the data flow of the 2D Poisson problem discussed in this work. 

\begin{table}[]
    \centering
    \begin{tabular}{ll}
Operation & Description\\
\hline \\
$(A_{m\times n} \circ B_{p \times q})$& Operator composition: $AB$ if $n = p$   \\
 $(A_{m\times n} \oplus B_{p \times q})$  & Direct sum operation: $\begin{bmatrix}
     A & \\
     & B
 \end{bmatrix}$ \\
 $(A_{m\times n} \otimes B_{p \times q})$ & Tensor product:\scriptsize
  $\begin{bmatrix} A_{00} B & \cdots & A_{0,n-1} B \\
  \vdots & \ddots & \vdots \\
  A_{m-1,0} B & \cdots & A_{m-1,n-1} B
  \end{bmatrix}$ \\
  $\begin{bmatrix}
      -\\
  \end{bmatrix}$ & Vertical stacking $\begin{bmatrix}
      \underline{A_{m\times n}}\\
      B_{p \times q}
  \end{bmatrix}$ \\
  $\begin{bmatrix}
      -\\
  \end{bmatrix}_{i = 0}^{k}$ & Iterative vertical stack \\
  $I_{n}$ & Identity matrix $I_{n \times n}$\\
  $pw_{x \mapsto f(x)}^{r \times s}$ & Pointwise operation \\
  $(a,b,c)$ & Row vector with three entries $a, b \text{ and } c$
\end{tabular}
    \caption{List of some operations used in SPIRAL for this work is shown here. The matrices $A_{m \times n}$ and $B_{p \times q}$ are considered as operators with $A: \mathbb{R}^{n} \rightarrow \mathbb{R}^{m}$ and $B:\mathbb{R}^{q} \rightarrow \mathbb{R}^{p} $.}
    \label{operators}
\end{table}

%One of the main difference between SPL and OL is that in OL the operators can have multiple input and output vectors. The mathematical operations described in OL are then placed into a rewrite system which in the end generates an optimized code for the problem specification. 

After expressing the given problem specification in terms of OL, the next step is to rewrite it in terms of $\Sigma-$OL. It is an intermediate step in SPIRAL and an extension of $\Sigma-$SPL \cite{Franchetti2005} which helps to incorporate loops into the problem specification unlike OL. This enables loop merging and as a result allows abstraction fusion. 

\section{ProtoX} \label{sec4}
\begin{figure}
    \centering
    \includegraphics[scale=0.7]{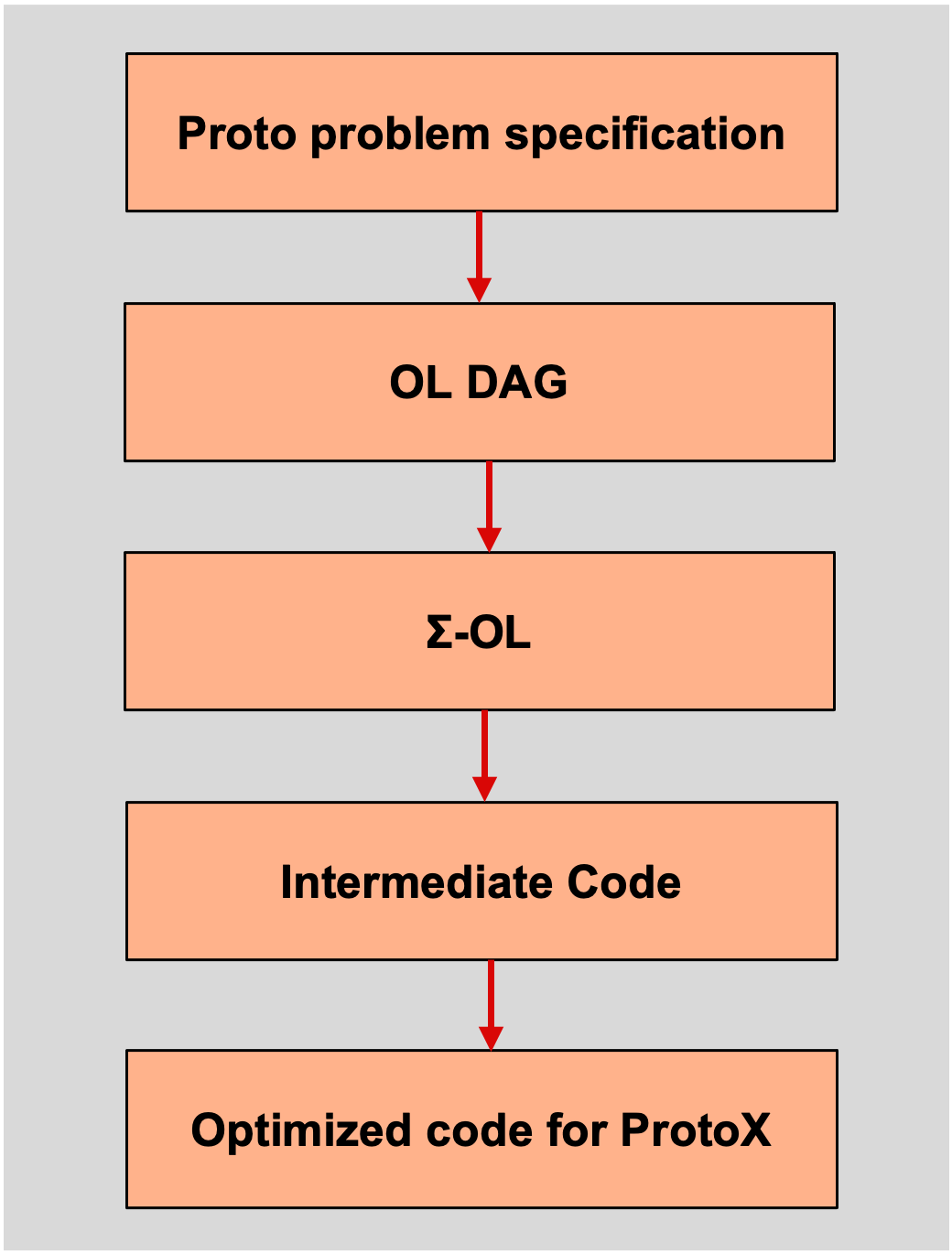}
    \caption{ProtoX design layout starting with a problem specification from Proto to the final optimized code generated using SPIRAL}
    \label{design}
\end{figure}

In this section we will describe the structure of ProtoX. The idea is to interpret Proto as a Domain Specific Language (DSL) with the help of SPIRAL. This is done by first interpreting an example from Proto as a mathematical specification and then map the Proto program specification to an OL expression. It is then broken down into a $\Sigma-$OL expression which introduces loop fusion. This will help generate a highly optimized C++ code. Here we will explain these ideas with respect to the 2D Poisson equation. The design layout for ProtoX is shown in Fig \ref{design}

{\bf{2D Poisson equation.}}
The Poisson equation is given as,
\begin{align} \label{Poisson}
    \Delta \phi(x,y) = \rho(x,y), \quad x,y \in \Omega := [0,1]\times [0,1],
\end{align}
where $\rho$ is a given function and $\phi$ is what we are solving for. $\Delta$ is the Laplace operator. We use the 5-pt stencil as a second order finite difference approximation of the Laplacian. The Jacobi iteration method is implemented in Proto to find the solution of (\ref{Poisson}). 
Let $h$ denote the mesh spacing for the discretized domain $\Omega$, then the Jacobi formula for a single iteration $n$ is given as
\begin{align}\label{jacobi}
    \phi_{i}^{n} := \phi_{i}^{n-1} + \lambda(S(\phi)_i^{n-1} -  \rho), 
\end{align}
where $\lambda = {h^2}/{4D}$ and $S(\phi)_{i} = \sum \limits_{{s} \in \mathbb{Z}^2} a_{s} 
\phi_{{i} + {s}}$ is the 5-point stencil applied to the input data $\phi_{i}^{n-1}$.

{\bf{Algorithm.}} In Proto, the domain space $\Omega$ is divided into several boxes and (\ref{Poisson}) is solved for each point in the box and then information is exchanged between the boxes to update the corresponding box data. The three main steps involved in this algorithm are 
\begin{enumerate}
    \item Applying the 5-pt Laplacian stencil to the given initial guess for $\phi$.
    \item Approximate the new value for $\phi$ using the Jacobi iteration method shown in (\ref{jacobi}). 
    \item Check the latest approximation of $\phi$ against the convergence criterion. Max norm is used in Proto to check for convergence.
\end{enumerate}
\begin{figure}
    \centering
    \includegraphics[scale=0.6]{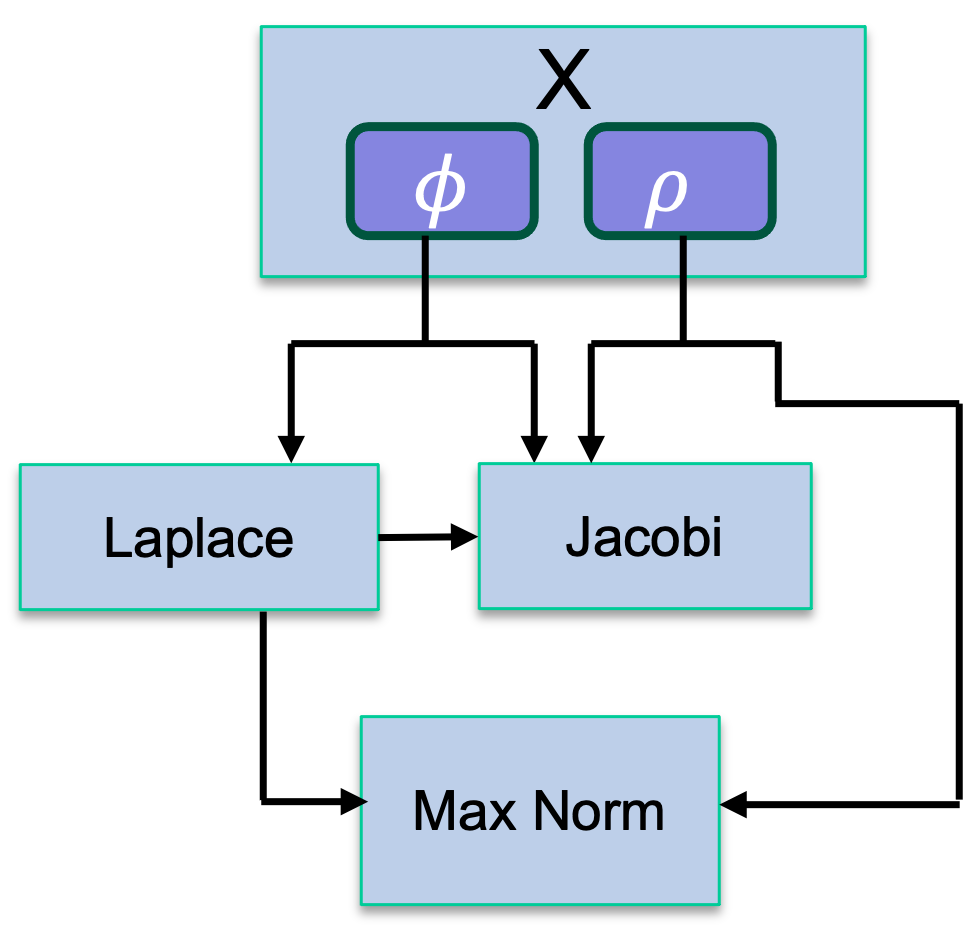}
    \caption{DAG for the Poisson problem in Proto}
    \label{DAG}
\end{figure}
We would like to note that in Proto each of the steps described in the algorithm above correspond to separate C++ function calls.  Figure \ref{Proto} provides the code sample that is used in Proto to solve (\ref{Poisson}). We can observe that the code reads in the same way as discussed in the algorithm above indicating an ease of use from a user perspective. However, this leads to intermediate data holders generating too much memory traffic. Hence, some optimization techniques need to be applied either at a compiler level or at an algorithmic level to overcome this issue. 
\begin{figure}[h]
 \begin{minted}[xleftmargin=15pt,linenos,fontsize=\scriptsize]{C}
// Defining the 5-pt Laplacian stencil
Stencil<double> laplace = Stencil<double>::Laplacian();
for (int iter = 0; iter < maxiter; iter++){
...
  // Solve for all boxes with each Box of size 64x64
  for (auto dit=phi.begin();*dit != dit.end();++dit){
    BoxData<double>& phiPatch = phi[*dit];
    BoxData<double>& rhoPatch = rho[*dit];

    // Compute the Laplacian
    BoxData<double> temp = laplace(phiPatch,wgt);

    // Jacobi iteration
    forallInPlace(jacobiUpdate,phiPatch,temp,
    rhoPatch,lambda);
 }
  // Computing || ||_{inf}
  double resmax=computeMaxResidualAcrossProcs(phi,
  rho,dx);
}
  \end{minted}
  \caption{Sample Proto code for the 2D Poisson problem}
  \label{Proto}
  \vspace{-1mm}
\end{figure}

{\bf{Implementation in SPIRAL. }} Figure \ref{DAG} provides the data flow for the algorithm discussed above. The first step in generating an optimized code using SPIRAL is to understand the data flow of the problem specification. Consequently, this data flow needs to be translated into OL with the help of the different operations shown in table \ref{operators}. 

We will consider the case where the size of each \mytt{Box} in Proto for this problem specification is $n \times n$ with $m \times m$ total elements in the box including ghost cells. The resulting OL expression corresponding to that data flow is shown below.  

\begin{align}
    \text{Poisson}_{n,m,t}^{\ell,w,a} &\rightarrow \begin{bmatrix}\scriptsize
        \text{Jacobi}_{n,m,w,l} \\
        \|.\|_\infty^{n, m, a} %\text{Maxnorm}_{n,m,a}
    \end{bmatrix} \circ \left(\begin{bmatrix}\scriptsize
            \text{I}_{n^2}  \\\scriptsize
         \text{Laplace} 
     \end{bmatrix}\oplus \text{I}_{n^2}\right) \circ X, \label{poissonOL}  \\
     \text{Laplace}_{n, m, t} &\rightarrow \text{Scatter}_{n^2 \times m^2} \circ [\text{Filt}(t)]_{i = 0}^{m^2}, \\ \label{laplaceOL}
     \text{Jacobi}_{n, m, w, l} &\rightarrow (1, w, -\lambda) \otimes \text{I}_{n^2}, \\ 
     \|.\|_\infty^{n, m, a} &\rightarrow \text{Max}\circ pw_{x \mapsto |x|}^{n \times n} \circ (0, 1/(a^2), -1) \otimes \text{I}_{n^2}. \label{normOL}
\end{align}
Here $X$ denotes the linearized input vector. For this problem specification it contains the initial data for $\phi \text{ and } \rho$ with $\phi$ being of size $m \times m \text{ or } m^2$ and $\rho$ is of size $n^2$. Hence, X is of size $n^2+m^2$. $t$ denotes the filter taps corresponding to the 5-pt stencil. Its a $3 \times 3$ matrix with entries as $[0,1,0], [1, -4, 1] \text{ and } [0,1,0]$ for the first, second and third row respectively. This matrix is flattened out and is iteratively applied to the input vector with the proper shifts. Scatter$_{n^2 \times m^2}$ denotes the \textit{scatter matrix} \cite{SpiralMultigrid} in SPIRAL. $\ell, w \text{ and }, a$ are scalar parameters used for the Jacobi iteration and the Laplacian.

The next step is to rewrite the OL breakdown rules shown in (\ref{poissonOL})-(\ref{normOL}), in terms of $\Sigma -$OL. This is where loop merging is introduced which helps in fusing different Proto abstractions into one single loop resulting in a considerable amount of performance gain. Here the computation is done per point in the data set rather than the entire patch which enables performing several computations simultaneously. A sample of the $\Sigma-$OL expression for this problem specification is shown below.
\begin{equation}
    \begin{split}
        \scriptsize \left(\sum_{j = 0}^{N-1} S_{r_j} \text{MaxNorm } G_{s_j} \right) \circ \left(\sum_{j = 0}^{N-1} S_{p_j} \text{Jacobi } G_{q_j} \right)\\
        \scriptsize \circ \left(\sum_{j = 0}^{N-1} S_{u_j} \text{Laplace } G_{t_j} \right).   
    \end{split}        
\end{equation}
Here $S$ and $G$ are the scatter and gather functions that are used in SPIRAL to read and write data \cite{Franchetti2005}. The subscripts $t_j, u_j, q_j, p_j, s_j \text{ and } r_j$ are the functions that indicate the exact number of points including their location that are being used to gather or read for computing a particular kernel (like the Laplace, Jacobi and MaxNorm in this case) and the number of points that are being scattered or written after the computation on the kernel is done.   

{\bf{Results.}}
Once the $\Sigma-$OL representation is achieved, an intermediate code is generated which in the end generates the final optimized C/C++ code for the given problem specification. Figures \ref{ProtoX}, \ref{ProtoXomp} and \ref{ProtoXgpu} provide the sample code for the 2D Poisson problem that is generated using SPIRAL. The code is generated for the \mytt{Box} size of $64 \times 64$ with the overall domain size being $256 \times 256$. We can see that in comparison to the original code in Proto (see Fig. \ref{Proto}) which has multiple abtractions for the different computational kernels required to solve the Poisson problem, the SPIRAL generated code only uses one abstraction and one loop. All the three kernels namely Laplace, Jacobi and MaxNorm are fused into one single loop. As a result this boosts the overall performance of the problem. 

\begin{figure}
    \centering
    \includegraphics[scale=0.6]{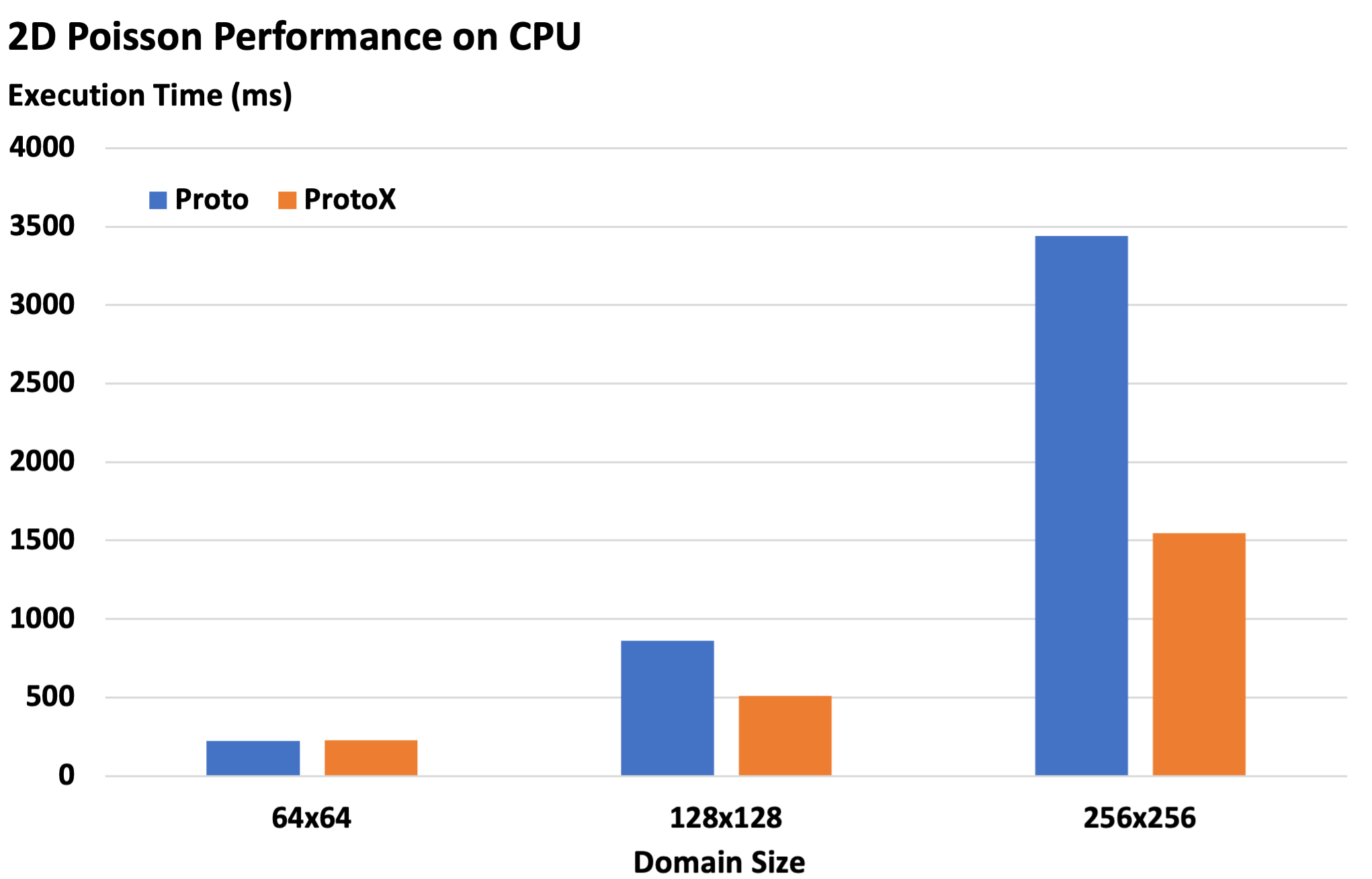}
    \caption{Run time comparison between the reference Proto code and the SPIRAL backed ProtoX code generated code for CPU. Here we are comparing different \mytt{Box} sizes ranging from $64\times64$ to $256 \times 256$ for a fixed $100$ iterations. We can observe that the ProtoX is performing up to $2 \times$ faster than the base Proto code. }
    \label{runtime}
\end{figure}

In Fig. \ref{runtime}, we provide a comparison of the run time between Proto and ProtoX. We compare three different box sizes - $64 \times 64, 128 \times 128 \text { and } 256 \times 256$. We keep the number of boxes fixed to be $4 \times 4$. This makes the corresponding domain sizes of the problem as $256 \times 256, 512 \times 512 \text{ and } 1024 \times 1024$ respectively. We ran both the Proto as well as the ProtoX code for a fixed $100$ Jacobi iterations. We can observe from the graph that ProtoX performes up to $2\times$ faster than the base Proto code for the 2D Poisson problem. We would like to remark here that these results were obtained on a local CPU machine with $2.3$GHz Quad-core Intel i7 processor. 

\begin{figure}
    \centering
    \includegraphics[scale=0.6]{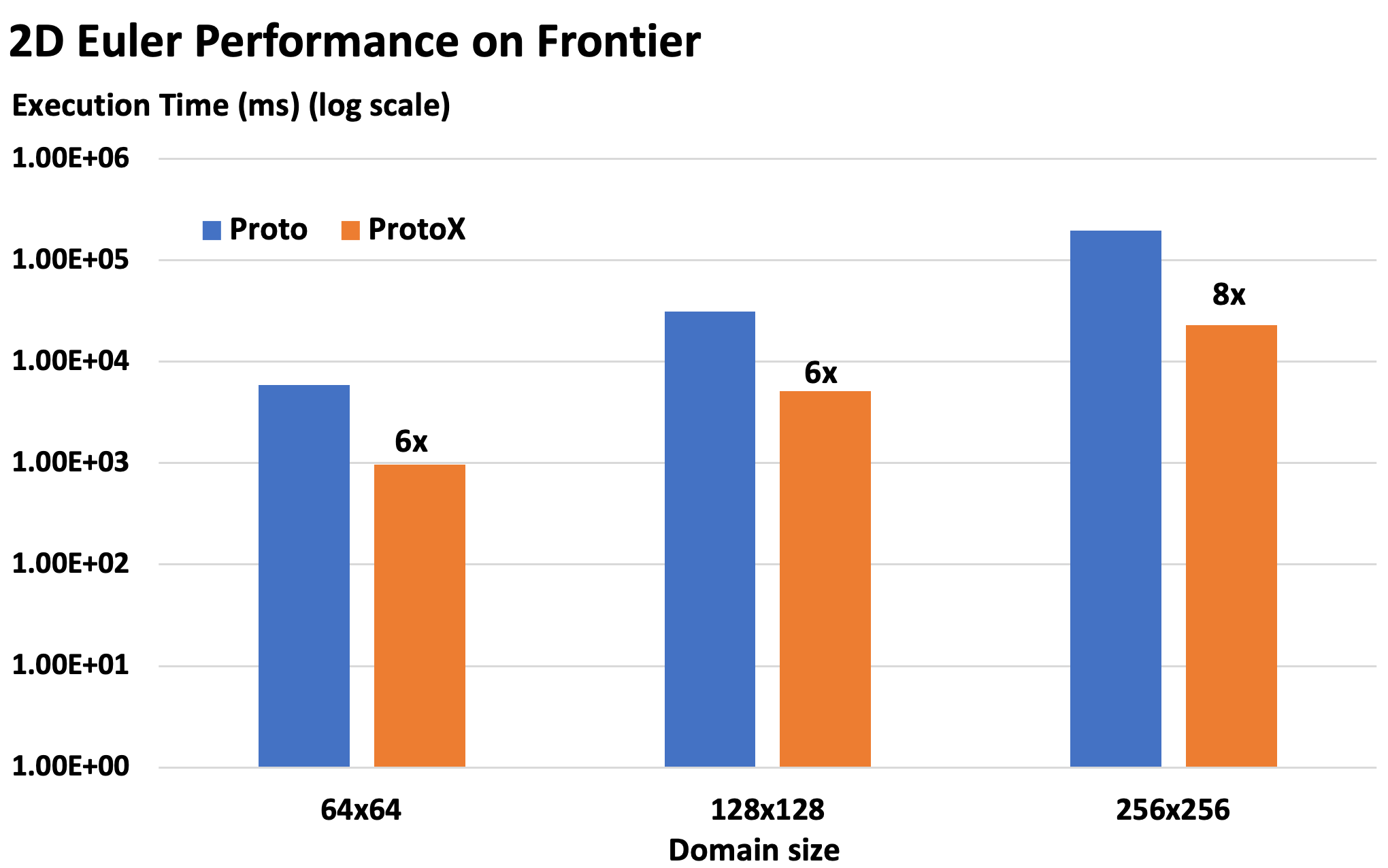}
    \caption{Initial run time results for 2D Euler equations. This is part of the ongoing work of adding more examples to ProtoX from the current Proto library. We can observe here that SPIRAL generated fused code for the Euler equations present in ProtoX provides up to $8 \times$ better performance compared to the baseline Proto code for the same problem.}
    \label{Euler}
\end{figure}

In addition to working on code generation for different targets, we are also working on adding more examples to ProtoX from the current Proto library. We are currently working on the Euler equations that are used in gas dynamics. Here finite volume method is used to solve these equations. This problem is much more complex and intricate with many different types of stencil and pointwise operations required to solve it. This provides a lot of room for optimization for SPIRAL which in return gives a significant improvement in the overall performance of the code.  Some of the initial CPU results are shown in Fig \ref{Euler}. We obtain up to $8 \times$ speedup over the base Proto code for the 2D Euler equation problem.

\begin{figure}[h]
 \begin{minted}[xleftmargin=15pt,linenos,fontsize=\scriptsize]{C}
void Poisson_2D_fused(double  *Y, double  *X,
    double weight1, double lambda1, double *rhs,
    double a_h1, double  *retval1){
        for(int i1 = 0; i1 <= 4095; i1++) {
            double s20, s21, s22;
            int a48, b15;
            b15 = ((66*(i1 / 64)) + (i1 % 64));
            a48 = (b15 + 67);
            s20 = X[a48];
            s21 = ((X[(b15 + 1)] - (4.0*s20)) 
            + X[(b15 + 66)] + X[(b15 + 68)]
            + X[(b15 + 133)]);
            s22 = rhs[i1];
            Y[a48] = ((s20 + (weight1*s21)) 
            - (lambda1*s22));
            *(retval1) = ((((*(retval1) >= 
            fabs((((1.0/(a_h1*a_h1))*s21)-s22)))))
            ? (*(retval1)) : 
            (fabs((((1.0/(a_h1*a_h1))*s21) 
            - s22))));
        }
}
  \end{minted}
  \caption{SPIRAL generated CPU code for the merged 2D Poisson equation for a \mytt{Box} of size $64 \times 64$. All the abstractions from Proto have been fused into one single function call and one single for loop}
  \label{ProtoX}
\end{figure}
\begin{figure}[h]
 \begin{minted}[xleftmargin=10pt,linenos,fontsize=\scriptsize]{C}
#include <omp.h>
#include <math.h>
const int NUM_THREADS = 4;

//SPIRAL code for OpenMP
void possion_2d(double  *Y, double  *X, double weight1, 
    double lambda1, double a_h1, double  *rhs, 
    double  *retval1) {
    #pragma omp parallel num_threads(4) 
    #reduction (max : retval)
    {
        { /* begin parallel loop */
            int tid1 = omp_get_thread_num(); 
            for(int i1 = tid1; i1 <= 4095; i1 += 4) {
                int a51, b15;
                double s28, s29, s30, s31, s32;
                b15 = ((66*(i1 / 64)) + (i1 % 64));
                a51 = (b15 + 67);
                s28 = X[a51];
                s29 = ((X[(b15 + 1)] - (((4.0)*(s28)))) + 
                      X[(b15 + 66)] + X[(b15 + 68)] + 
                      X[(b15 + 133)]);
                s30 = rhs[i1];
                s31 = ((s28 + ((weight1)*(s29))) - 
                      (((lambda1)*(s30))));
                Y[a51] = s31;
                s32 = max(*(retval1), 
                    abs((((((1.0) / (((a_h1)*(a_h1)))))
                    *(s29))- (s30))));
                *(retval1) = s32;
            }
        } /* end parallel loop */
    }
}
  \end{minted}
  \caption{SPIRAL generated code for OpenMP}
  \label{ProtoXomp}
\end{figure}
\begin{figure}[h]
 \begin{minted}[xleftmargin=10pt,linenos,fontsize=\scriptsize]{C}
#include "hip/hip_runtime.h"

__global__ void ker_code0(double  *X, double  *Y, 
    double weight1, double lambda1,double a_h1, 
    double  *retval1) {
    if (((((256*blockIdx.x) + threadIdx.x) < 4096))) {
        double s21, s22;
        int a66, a67, b16;
        a66 = (threadIdx.x + (256*blockIdx.x));
        b16 = ((66*(a66 / 64)) + (threadIdx.x % 64));
        a67 = (b16 + 67);
        s21 = X[a67];
        s22 = ((X[(b16 + 1)] - (4.0*s21)) + X[(b16 + 66)]
            + X[(b16 + 68)] + X[(b16 + 133)]);
        Y[a67] = ((s21 + (weight1*s22)) - (lambda1
                *X[(a66 + 4356)]));
        *(retval1) = ((((*(retval1) >= 
            fabs((((1.0 / (a_h1*a_h1))*s21) - s22))))) ? 
            (*(retval1)): 
            (fabs((((1.0 / (a_h1*a_h1))*s21) - s22))));
    }
}
void possion_2d(double  *Y, double  *X, double weight1,
        double lambda1, double a_h1, double  *rhs,
        double  *retval1) {
    dim3 b17(256, 1, 1), g1(17, 1, 1);
    hipLaunchKernelGGL(ker_code0, dim3(g1), dim3(b17),
    0, 0, X, Y, weight1, lambda1, a_h1, retval1);
}
  \end{minted}
  \caption{SPIRAL generated code for GPU}
  \label{ProtoXgpu}
\end{figure}

\section{Conclusion and Future Work}\label{sec5}
This work demonstrates a proof of concept of having SPIRAL code generation as a backend to Proto using the 2D Poisson problem. It shows that by writing a Proto program as a SPIRAL specification, we can interpret Proto as a DSL and powerful code generation is possible. This work is an early prototype, so only CPU performance for some specific sizes has been shown. Considering that these are initial results, we can still observe that there is a reasonable amount of performance improvement happening for the given problem specification. The fused code for GPU and multi-core CPU is also shown. The run time performance results for these two architectures is part of the ongoing work. 

 Proto is a full featured packaged library that is being used for many scientific applications like astrophysics, particle physics, fluid and gas dynamics and many more. It has many implementations of different types of numerical algorithms like the Adaptive Mesh Refinement (AMR), multigrid method and finite volume method. Hence, one of the future goals is to bring ProtoX at par with Proto in terms of the examples and different target platforms used in there. Other part of the future work is to make ProtoX interoperable with FFTX \cite{FFTX} to do cross library optimization.

\section{Acknowledgment}\label{sec6}
This project is funded by the  DOE Office of Advanced Scientific Computing Research the Office of Science of the U.S. Department of Energy under Contract No. DE-AC02-05CH11231.

% \section*{Acknowledgment}
\bibliographystyle{IEEEtran}
\bibliography{references}

% Generated by IEEEtran.bst, version: 1.14 (2015/08/26)
\begin{thebibliography}{10}
\providecommand{\url}[1]{#1}
\csname url@samestyle\endcsname
\providecommand{\newblock}{\relax}
\providecommand{\bibinfo}[2]{#2}
\providecommand{\BIBentrySTDinterwordspacing}{\spaceskip=0pt\relax}
\providecommand{\BIBentryALTinterwordstretchfactor}{4}
\providecommand{\BIBentryALTinterwordspacing}{\spaceskip=\fontdimen2\font plus
\BIBentryALTinterwordstretchfactor\fontdimen3\font minus
  \fontdimen4\font\relax}
\providecommand{\BIBforeignlanguage}[2]{{%
\expandafter\ifx\csname l@#1\endcsname\relax
\typeout{** WARNING: IEEEtran.bst: No hyphenation pattern has been}%
\typeout{** loaded for the language `#1'. Using the pattern for}%
\typeout{** the default language instead.}%
\else
\language=\csname l@#1\endcsname
\fi
#2}}
\providecommand{\BIBdecl}{\relax}
\BIBdecl

\bibitem{Spiral2005}
M.~Puschel, J.~Moura, J.~Johnson, D.~Padua, M.~Veloso, B.~Singer, J.~Xiong,
  F.~Franchetti, A.~Gacic, Y.~Voronenko, K.~Chen, R.~Johnson, and N.~Rizzolo,
  ``Spiral: Code generation for dsp transforms,'' \emph{Proceedings of the
  IEEE}, vol.~93, no.~2, pp. 232--275, 2005.

\bibitem{Franchetti2005}
F.~Franchetti, Y.~Voronenko, and M.~P\"{u}schel, ``Formal loop merging for
  signal transforms,'' \emph{SIGPLAN Not.}, vol.~40, no.~6, p. 315–326, jun
  2005.

\bibitem{FFTX}
F.~Franchetti, D.~G. Spampinato, A.~Kulkarni, D.~Thom~Popovici, T.~M. Low,
  M.~Franusich, A.~Canning, P.~McCorquodale, B.~V. Straalen, and P.~Colella,
  ``Fftx and spectralpack: A first look,'' in \emph{2018 IEEE 25th
  International Conference on High Performance Computing Workshops (HiPCW)},
  2018, pp. 18--27.

\bibitem{Sanil2020}
S.~Rao, A.~Kutuluru, P.~Brouwer, S.~McMillan, and F.~Franchetti, ``Gbtlx: A
  first look,'' in \emph{2020 IEEE High Performance Extreme Computing
  Conference (HPEC)}, 2020, pp. 1--7.

\bibitem{halide}
J.~Ragan-Kelley, C.~Barnes, A.~Adams, S.~Paris, F.~Durand, and S.~Amarasinghe,
  ``Halide: A language and compiler for optimizing parallelism, locality, and
  recomputation in image processing pipelines,'' vol.~48, no.~6, 2013.

\bibitem{LIFT}
B.~Hagedorn, L.~Stoltzfus, M.~Steuwer, S.~Gorlatch, and C.~Dubach, ``High
  performance stencil code generation with lift,'' ser. CGO 2018.\hskip 1em
  plus 0.5em minus 0.4em\relax New York, NY, USA: Association for Computing
  Machinery, 2018, p. 100–112.

\bibitem{SBLOCK}
T.~Brandvik and G.~Pullan, ``Sblock: A framework for efficient stencil-based
  pde solvers on multi-core platforms,'' in \emph{2010 10th IEEE International
  Conference on Computer and Information Technology}, 2010, pp. 1181--1188.

\bibitem{BVSCompiler2015}
P.~Basu, M.~Hall, S.~Williams, B.~Van~Straalen, L.~Oliker, and P.~Colella,
  ``Compiler-directed transformation for higher-order stencils,'' in \emph{2015
  IEEE International Parallel and Distributed Processing Symposium}, 2015, pp.
  313--323.

\bibitem{PATUS}
M.~Christen, O.~Schenk, and H.~Burkhart, ``Patus: A code generation and
  autotuning framework for parallel iterative stencil computations on modern
  microarchitectures,'' in \emph{2011 IEEE International Parallel and
  Distributed Processing Symposium}, 2011, pp. 676--687.

\bibitem{SadayGPU}
J.~Holewinski, L.-N. Pouchet, and P.~Sadayappan, ``High-performance code
  generation for stencil computations on gpu architectures,'' in
  \emph{Proceedings of the 26th ACM International Conference on
  Supercomputing}, ser. ICS '12.\hskip 1em plus 0.5em minus 0.4em\relax New
  York, NY, USA: Association for Computing Machinery, 2012, p. 311–320.

\bibitem{ManyCore}
M.~Li, Y.~Liu, H.~Yang, Y.~Hu, Q.~Sun, B.~Chen, X.~You, X.~Liu, Z.~Luan, and
  D.~Qian, ``Automatic code generation and optimization of large-scale stencil
  computation on many-core processors,'' in \emph{Proceedings of the 50th
  International Conference on Parallel Processing}, ser. ICPP '21.\hskip 1em
  plus 0.5em minus 0.4em\relax New York, NY, USA: Association for Computing
  Machinery, 2021.

\bibitem{ExaStencil}
C.~Lengauer, S.~Apel, M.~Bolten, A.~Gr{\"o}{\ss}linger, F.~Hannig,
  H.~K{\"o}stler, U.~R{\"u}de, J.~Teich, A.~Grebhahn, S.~Kronawitter,
  S.~Kuckuk, H.~Rittich, and C.~Schmitt, ``Exastencils: Advanced stencil-code
  engineering,'' in \emph{Euro-Par 2014: Parallel Processing Workshops},
  L.~Lopes, J.~{\v{Z}}ilinskas, A.~Costan, R.~G. Cascella, G.~Kecskemeti,
  E.~Jeannot, M.~Cannataro, L.~Ricci, S.~Benkner, S.~Petit, V.~Scarano,
  J.~Gracia, S.~Hunold, S.~L. Scott, S.~Lankes, C.~Lengauer, J.~Carretero,
  J.~Breitbart, and M.~Alexander, Eds.\hskip 1em plus 0.5em minus 0.4em\relax
  Cham: Springer International Publishing, 2014, pp. 553--564.

\bibitem{YASK}
C.~Yount, J.~Tobin, A.~Breuer, and A.~Duran, ``Yask—yet another stencil
  kernel: A framework for hpc stencil code-generation and tuning,'' in
  \emph{2016 Sixth International Workshop on Domain-Specific Languages and
  High-Level Frameworks for High Performance Computing (WOLFHPC)}, 2016, pp.
  30--39.

\bibitem{InPlaceStencil}
M.~Essadki, B.~Michel, B.~Maugars, O.~Zinenko, N.~Vasilache, and A.~Cohen,
  ``Code generation for in-place stencils,'' in \emph{Proceedings of the 21st
  ACM/IEEE International Symposium on Code Generation and Optimization}, ser.
  CGO 2023.\hskip 1em plus 0.5em minus 0.4em\relax New York, NY, USA:
  Association for Computing Machinery, 2023, p. 2–13.

\bibitem{SPL1}
J.~Xiong, J.~Johnson, R.~Johnson, and D.~Padua, ``Spl: A language and compiler
  for dsp algorithms,'' \emph{SIGPLAN Not.}, vol.~36, no.~5, p. 298–308, may
  2001.

\bibitem{SpiralOL2009}
F.~Franchetti, F.~de~Mesmay, D.~McFarlin, and M.~P{\"u}schel, ``Operator
  language: A program generation framework for fast kernels,'' in
  \emph{Domain-Specific Languages}, W.~M. Taha, Ed.\hskip 1em plus 0.5em minus
  0.4em\relax Berlin, Heidelberg: Springer Berlin Heidelberg, 2009, pp.
  385--409.

\bibitem{SpiralMultigrid}
M.~Bolten, F.~Franchetti, P.~H.~J. Kelly, C.~Lengauer, and M.~Mohr, ``Algebraic
  description and automatic generation of multigrid methods in spiral,''
  \emph{Concurrency and Computation: Practice and Experience}, vol.~29, no.~17,
  p. e4105, 2017, e4105 cpe.4105.

\end{thebibliography}

\end{document}